\documentclass[
 reprint,
superscriptaddress,
 amsmath,amssymb,
 aps,
]{revtex4-2}

\usepackage{graphicx}
\usepackage{dcolumn}
\usepackage{bm}

\usepackage{xcolor}
\usepackage{hyperref}

\def\nue{\mathrel{{\nu_e}}}

\def\barnue{\mathrel{{\bar \nu}_e}}

\def\msun{\mathrel{{M_{\odot} }}}
\newcommand{\n}{neutrino}

\newcommand{\gwb}{SGWB}
\newcommand{\sngwb}{SN-SGWB}

\newcommand{\be}{\begin{equation}}
\newcommand{\ee}{\end{equation}}
\newcommand{\ba}{\begin{eqnarray}}
\newcommand{\ea}{\end{eqnarray}}

\begin{document}

\preprint{APS/123-QED}

\title{A ``Neutrino Fog" For Gravitational Waves: the Stochastic Gravitational Wave Background from Supernova Neutrino Memory}

\author{Alex Rojewski}

\author{Cecilia Lunardini}%

\affiliation{%
 Department of Physics, Arizona State University\\
 450 E. Tyler Mall, Tempe, AZ 85287-1504 USA
}%

\date{August 4, 2026}

\begin{abstract}
Gravitational waves originating from unresolved sources, or stochastic gravitational wave backgrounds (SGWBs), carry precious information about the physics underlying their diverse sources, and represent an important target for future experimental searches. Here, we model the SGWB due to core collapse supernovae.  Using an extensive collection of state-of-the-art, three-dimensional, multi-second supernova simulations, we characterize the two main components of this background:  one at $f \sim 10^{-2} - 1$ Hz, due to anisotropic neutrino emission (the gravitational wave memory); the other at $f \gtrsim  10^{2} $ Hz, from the near-core matter dynamics. 
We find that the memory component offers the best prospects of detection, as its characteristic peak at $f\sim 0.1$ Hz is within the reach of future space-born detectors. At the peak, the energy density might be comparable to that of backgrounds from slow roll inflation and from possible cosmological relics, thus potentially impacting searches of these important signals. 

\end{abstract}

\maketitle

\section{\label{sec:level1}Introduction} \label{introduction}

Since the first detection of gravitational waves (GWs) in 2015 \cite{LIGOScientific:2016aoc}, gravitational wave astronomy has become an important addition to the field of multi-messenger astronomy. GW observations of compact binary mergers have provided a wealth of information about the individual merging objects, merger rates, a possible pair-instability mass gap, and have been used as test of general relativity \cite{LIGOScientific:2017vwq,LIGOScientific:2020aai,LIGOScientific:2020iuh,LIGOScientific:2020ibl,KAGRA:2021vkt,LIGOScientific:2021qlt,LIGOScientific:2025rsn,LIGOScientific:2020tif,LIGOScientific:2025rid,LIGOScientific:2025wao,Tong:2025wpz}.  It is expected that, as the field matures, GWs from other classes of sources, such as binary supermassive black holes, core collapse supernovae, and the physics of the early Universe, will eventually be observed.

The discovery of GW transients 
implies the existence of a stochastic gravitational wave background (SGWB), an incoherent GW signal produced by populations of unresolved sources \cite{Michelson:1987,Christensen:1992,Flanagan:1993,Allen:1997ad,Regimbau:2008nj,Regimbau:2011rp,Christensen:2018iqi,Caprini:2018mtu,Renzini:2022alw}. Prior to 2023, all observed GW signals came from individually resolved events, and to date the SGWB from stellar-mass compact binary mergers (or, coalescences, CBCs) has not been detected \cite{LIGOScientific:2016fpe,KAGRA:2021kbb}.
In 2023, several pulsar timing arrays (PTAs) detected a signal consistent with a SGWB, thought to originate from a background of unresolved supermassive black hole mergers \cite{NANOGrav:2023gor,EPTA:2023fyk,Reardon:2023gzh,Xu:2023wog}. These findings have reinvigorated the interest in SGWBs, both from CBCs \cite{Christensen:2018iqi,Regimbau:2008nj,Regimbau:2011rp,KAGRA:2021kbb} and from other astrophysical \cite{Regimbau:2011rp,Christensen:2018iqi,Buonanno:2004tp,Crocker:2017agi,Finkel:2021zgf,Marassi:2010wj,Owen:1998xg,Rosado:2012bk,Cheng:2017kmv} or cosmological sources \cite{Caprini:2018mtu,Christensen:2018iqi,Renzini:2022alw,LISACosmologyWorkingGroup:2022jok,Guzzetti:2016mkm,Barnaby:2011qe,Turner:1996ck,Bartolo:2016ami,Blanco-Pillado:2017rnf,Auclair:2019wcv,Avgoustidis:2025svu,Caprini:2015zlo,Schmitz:2020syl,Croon:2024mde,LIGOScientific:2025kry}.

In this work, we explore the contribution of core collapse supernovae to the \gwb\ (\sngwb). The \sngwb\ was first estimated in the seminal study in Ref. \cite{Buonanno:2004tp}, where its two main components were discussed: (i) a higher frequency one ($f\sim 100 $ Hz) originating from the newly-formed proto-neutron star (PNS) or from hydrodynamic effects within the shock front, and (ii) a lower frequency term due to the contributions from the gravitational wave memory effect from escaping supernova neutrinos \cite{Zeldovich:1974gvh,Turner:1978jj,Epstein:1978dv, Braginsky:1987kwo,Mukhopadhyay:2021zbt}. 
Subsequent studies  \cite{Crocker:2017agi, Finkel:2021zgf, Chowdhury:2024fdr}, focused on better characterizing the former, with relatively little attention given to the latter.  They were driven by the rapid progress in numerical modeling of the  waveforms in the 10-100 Hz band, the success of second-generation GW detectors, and plans for the construction of third-generation detectors with greatly improved sensitivities in the same frequency band \cite{Evans:2023euw,ET:2025xjr}.

Improvements in the modeling of the neutrino memory component have required a longer time to develop, due to the key role of the asymmetry of the neutrino emission, which requires fully three dimensional (3D), multi-second simulations to be properly characterized.
Recently, these goals have been achieved by several groups, resulting in an outpouring of new results \cite{Bollig:2020phc,Burrows:2024pur,Choi:2024irp,Choi:2025igp,Lella:2026aat} with progress also being made towards inclusion of rotation and magnetic fields \cite{Shibagaki:2023tmh,Nakamura:2024qhx,Sykes:2024mel,Powell:2024nvv}. In addition, recent work has suggested that the neutrino memory component from a single supernova may be detectable at distances up to $10$ kpc by third-generation detectors \cite{Vartanyan:2020nmt,Mukhopadhyay:2021zbt,Richardson:2021lib,Shibagaki:2023tmh}, and possibly to $100$ kpc with futuristic detectors and favorable progenitor properties \cite{Powell:2024nvv}.

Motivated by these advances, here we present a modern study of the \sngwb, with emphasis on its neutrino memory component.  
We model this component using results from state-of-the-art, long-duration 3D simulations from \cite{Burrows:2024pur,Choi:2024irp,Choi:2025igp}. We also include estimates of the signals from PNS and hydrodynamic contributions for  comparison. For both memory and matter components, we account for the effects of varying progenitor masses, and discuss how the \sngwb\ compares to stochastic backgrounds from other classes of sources, especially in view of upcoming next-generation gravitational wave detectors with sub-Hz sensitivity.

This paper is organized as follows. In Section \ref{formalism}, the relevant formalism is presented. In section \ref{sec:level1}, we introduce the numerical simulations adopted here, and present results for our fits to the individual waveforms for memory and matter contributions. 
In section \ref{sec:level2}, we present results for the SGWB, detailing the effects of integrating over the stellar population. 
Observational prospects are presented as well. In section \ref{sec:discussion}, we outline possible next steps and discuss our results in the broader context of the field, with emphasis on the possible role of the stochastic background from memory in searches for gravitational waves from inflation and cosmological relics. Appendix \ref{inputVariation} discusses the effect of varying source orientations and of uncertainties in the cosmic star formation rate. 

\section{Formalism} \label{formalism}

\subsection{Stochastic Gravitational Wave Background}

The dimensionless energy density spectrum of gravitational waves is defined as \cite{Allen:1997ad,Regimbau:2008nj,Regimbau:2011rp,LIGOScientific:2016fpe}: 
\begin{equation}
\Omega_{\text{GW}}\left(f\right)=\frac{1}{\rho_{c}}\frac{d\rho_{\text{GW}}}{d\ln{f}},
\end{equation}
where $d\rho_{\text{GW}}$ is the energy density of gravitational waves in the frequency interval $\left(f, f+df\right)$ and $\rho_{c}=3H^{2}_{0}c^{2}/8\pi G$ is the critical energy density required to close the universe. For astrophysical sources, we may write $d\rho_{\text{GW}}/d\ln{f}$ as an integral over redshift \cite{Phinney:2001di,Buonanno:2004tp,LIGOScientific:2016fpe, Crocker:2017agi}, such that, for a population of identical sources, 
\begin{eqnarray}
\Omega_{\text{GW}}
\left(f\right)
&=&\frac{f}{\rho_{c}H_{0}} \nonumber \\
&\times &\int_{0}^{z_{max}}\frac{ dz R\left(z\right)}{\left(1+z\right)\mathcal{E}\left(\Omega_{m},\Omega_{\Lambda},z\right)} \frac{dE}{df_{s}}\left(f_{s}\right). 
\label{eq:masterintegral}
\end{eqnarray}
Here, $f_{s}$ is the frequency at the source, which is related to the observed frequency $f$ by $f_{s}=f\left(1+z\right)$. $z_{max}$ is the maximum redshift at which there exist sources that contribute to the stochastic background; we chose the value $z_{max}=30$, as this is the approximate redshift at which the first stars formed (see, e.g. \cite{Klessen:2023qmc}). The $\left(1+z\right)$ factor in the denominator accounts for cosmic expansion by converting time in the source frame to the detector frame \cite{LIGOScientific:2016fpe}. The product $H_{0} \mathcal{E}\left(\Omega_{m},\Omega_{\Lambda},z\right)$ accounts for the redshift-dependence of the comoving volume, where  $\mathcal{E}\left(\Omega_{m},\Omega_{\Lambda},z\right)$ is
\begin{equation}
\mathcal{E}\left(\Omega_{m},\Omega_{\Lambda},z\right)=\sqrt{\Omega_{m}\left(1+z\right)^{3}+\Omega_{\Lambda}}~.
\end{equation}
We assume a flat universe and ignore the radiation energy density contribution, which is negligible in the redshift interval of interest. We also assume a cosmology consistent with the \textit{Planck} 2018 results: $\Omega_{m}=0.3111$, $\Omega_{\Lambda}=0.6889$, and $H_0=67.66$ km s$^{-1}$ Mpc$^{-1}$ \cite{Planck:2018vyg}.

The function $R\left(z\right)$ encodes the rate of GW-producing events per unit comoving volume per unit source time  \cite{LIGOScientific:2016fpe}. For our purposes, this is the rate of core collapse supernovae.   $R\left(z\right)$ is proportional to the star formation rate  (SFR), $R_{*}\left(z\right)$ (see, e.g. \cite{Crocker:2017agi,Finkel:2021zgf,Chowdhury:2024fdr}):
\begin{equation}\label{eq:CCR}
R\left(z\right)=\lambda_{CC}R_{*}\left(z\right)~,
\end{equation}
with $\lambda_{CC} \approx 0.007M_{\odot}^{-1}$ being the mass fraction of stars that undergo core collapse, which we estimate assuming the Salpeter initial mass function (IMF) and a progenitor star mass range $M \geq 8~M_\odot$.

In our fiducial model, we parameterize the SFR as in Ref. \cite{Hernquist:2002rg}:
\begin{equation}\label{eq:SFR}
R_{*}\left(z\right)=\nu\frac{pe^{q\left(z-z_{m}\right)}}{p-q+qe^{p\left(z-z_{m}\right)}}~,
\end{equation}
with parameter values from Ref. \cite{Vangioni:2014axa}:  $\nu=0.178$ $M_{\odot}$/yr/Mpc$^{3}$, $p=2.37$, $q=1.80$, and $z_m = 2.00$. This model is based on the observed luminosity function of high-redshift galaxies, and is consistent with global metallicity observations \cite{Vangioni:2014axa}. 

In Eq. (\ref{eq:masterintegral}), $\frac{dE}{df_{s}}\left(f_{s}\right)$ is the gravitational wave energy spectrum due to a single core-collapse supernova.  We may write this as \cite{Phinney:2001di,Buonanno:2004tp, Crocker:2017agi, Finkel:2021zgf}: 
\begin{equation} \label{dedf}
\frac{dE}{df_{s}}\left(f_{s}\right)=\frac{\pi^{2}c^{3}r^{2}}{G}f_{s}^{2}\left\langle\left|\tilde{h}_{+}\left(f_{s}\right)\right|^{2}+\left|\tilde{h}_{\times}\left(f_{s}\right)\right|^{2}\right\rangle_{\Omega},
\end{equation}
where $r$ is the distance to the source (in our examples,  $r=10$ kpc will be used), and $\left\langle\cdot\right\rangle_{\Omega}$ indicates that we take the average over source orientations \cite{Phinney:2001di}.  Here $\tilde{h}_{+}$ and $\tilde{h}_{\times}$ are the Fourier transforms of the plus and cross mode strains, respectively, of the gravitational wave signal.

\subsection{Gravitational Wave Memory}

In addition to the more familiar oscillatory waveform  where the strain starts at and ultimately returns to zero, gravitational waves can also have a component that asymptotes to a non-zero strain, the gravitational wave memory \cite{Zeldovich:1974gvh,Turner:1978jj,Epstein:1978dv,Braginsky:1987kwo,Thorne:1992sdb,Blanchet:1992br,Favata:2010zu}. This component is present if (1) part of the system is (or becomes) gravitationally unbound, and (2) there is anisotropy in the unbound mass or radiation \cite{Favata:2010zu}. For the case of a supernova neutrino burst, the strain can be parameterized as \cite{Epstein:1978dv}:
\begin{equation} \label{numemory}
h_{i}\left(t,\mathbf{\Omega}\right)=\frac{2G}{rc^{4}}\int^{t-r/c}_{-\infty}dt' L_{\nu}\left(t'\right)\alpha_{i}\left(t',\mathbf{\Omega}\right),
\end{equation}
where $L_{\nu}\left(t'\right)$ is the total luminosity in neutrinos emitted by the source, and $\alpha_{i}\left(t',\mathbf{\Omega}\right)$ is the anisotropy parameter (which may be a function of source orientation, $\mathbf{\Omega}$, with respect to the observer) that encodes the deviation from spherical symmetry. The index $i\in \left\{+,\times\right\}$ indicates the gravitational wave mode. 

Here we adopt a convenient phenomenological model inspired by the ``wl4GNZ'' model in Ref. \cite{Mukhopadhyay:2021zbt}. It consists of modeling  $L_{\nu}(t)$ as a decaying exponential (which is a good approximation for the first $\sim 10$ s or so of the burst),
\begin{equation} \label{nuluminosity}
L_{\nu}\left(t\right)=\beta e^{-\chi t}~,
\end{equation}
and $\alpha_{i}$ as a sum of Gaussians,
\begin{equation} \label{nuanisotropy}
\alpha_{i}\left(t,\mathbf{\Omega}\right)=\sum_{j=1}^{N}\xi_{j}\exp{\left(-\frac{\left(t-\gamma_{j}\right)^{2}}{2\sigma_{j}^{2}}\right)}~.
\end{equation}
We note that the amplitude of the memory waveform depends on the orientation of the source with respect to the observer, $\mathbf{\Omega}$; the dependence enters the calculation through the  $\xi_{j}$ parameters.

The parameterizations in Eqs. (\ref{nuluminosity}) and (\ref{nuanisotropy}) lead to simple forms for the strain in the time and frequency domain, as follows \cite{Mukhopadhyay:2021zbt}: 
\begin{equation} \label{nuphenostrain}
h\left(t,\mathbf{\Omega}\right)=\sum_{j=1}^{N}\kappa_{j}\left[\text{erf}\left(\zeta_{j}\tau_{j}\right)+\text{erf}\left(\zeta_{j}\left(t-\tau_{j}\right)\right)\right]~; 
\end{equation}
\begin{equation} \label{numemspectrum}
\tilde{h}\left(f,\mathbf{\Omega}\right)=\sum_{j=1}^{N}\kappa_{j}\frac{i}{\pi f}\exp{\left(-\frac{\pi^{2}f^{2}}{\zeta_{j}^{2}}\right)}\exp{\left(i2\pi f\tau_{j}\right)}~,
\end{equation}
where we used the convention:
\begin{eqnarray}
\tilde{h}\left(f\right)=\int_{-\infty}^{\infty} h\left(t\right)\text{ }e^{2\pi ift}\text{ }dt,
\end{eqnarray}
and the quantities $\kappa_j, \zeta_j, \tau_j$ are:
\begin{eqnarray}
\kappa_{j}&&=\frac{2G}{rc^{4}}\sqrt{\frac{\pi}{2}}\beta\xi_{j}\sigma_{j}\exp{\left(\frac{\chi}{2}\left(-2\gamma_{j}+\sigma_{j}^{2}\chi\right)\right)},\nonumber\\
\zeta_{j}&&=\frac{1}{\sqrt{2}\sigma_{j}},\nonumber\\
\tau_{j}&&=\gamma_{j}-\sigma_{j}^{2}\chi.\nonumber\\
\end{eqnarray}
While we have now suppressed the mode index on the strain $h\left(t,\mathbf{\Omega}\right)$, we note that we will have separate parameter sets $\left\{\kappa_{j},\zeta_{j},\tau_{j}\right\}$ for each mode. We also note that since the $\xi_{j}$ carry the dependence on the source orientation, the coefficients $\kappa_j$ are likewise orientation-dependent.

For simplicity, from here on the angular argument $\mathbf{\Omega}$ will be omitted, since only fixed, discrete values for it will be used (see Sec. \ref{sims}).

\subsection{Matter Contribution}

In addition to the memory, we model the contribution of matter to the \sngwb\ in a consistent manner. Gravitational wave emission from matter motion in CCSNe arises primarily from excitations of oscillatory modes in the PNS by accreting matter \cite{Radice:2018usf,Szczepanczyk:2021bka,Vartanyan:2023sxm,Muller:2026ofz}, with a lower-frequency component arising from hydrodynamical instabilities and convection in the layers between the PNS and the shock \cite{Blondin:2002sm,Fernandez:2010db,Muller:2026ofz}.
A useful description  of it is the combined low- and high-frequency phenomenological model in Ref. \cite{Crocker:2017agi}, where terms modeling the signals from PNS oscillations and from hydrodynamic effects are included. 
We use a modified version of this model, which reads as follows: 
\begin{eqnarray} \label{matterpheno}
&&\frac{dE}{df_{s}}
=\frac{\zeta}{\lambda_{CC}}\left[\pi\sqrt{\frac{c^{3}}{\zeta G}}\left(A'\exp{\left(-\frac{\left(f_{s}-\mu_{1}\right)^{2}}{2\omega_{1}^{2}}\right)}\right.\right.\nonumber\\
&& \left.\left.-B'\exp{\left(-\frac{\left(f_{s}-\mu_{2}\right)^{2}}{2\omega_{2}^{2}}\right)}\right) 
+\left(1+\frac{f_{s}}{a}\right)^{3}e^{-\frac{f_{e}}{b}}\right]^{2}~.
\end{eqnarray}
In addition to the two terms in \cite{Crocker:2017agi}, we add a Gaussian with negative amplitude ($-B^\prime<0$), that models the suppression in the kilohertz band due to an avoided \textit{f-g} mode crossing \cite{Aizenman:1976,Sotani:2020eva,Vartanyan:2023sxm}. In some cases $B^\prime=0$ will be taken, if such a feature is invisible in the simulation results we use.

\begin{table}[h!]
\centering
\begin{tabular}{|c c c c|} 
 \hline
 Progenitor & & Compact & \\
 Mass $\left(M_{\odot}\right)$ & Duration $\left(\text{s}\right)$ & Remnant & Exploding? \\ [0.5ex] 
 \hline
 9 & 2.01 & NS & Y \\ 
 9.25 & 2.75 & NS & Y \\
 9.5 & 2.14 & NS & Y \\
 11 & 3.08 & NS & Y \\
 12.25 & 2.01 & BH & N \\ 
 14 & 2.49 & BH & N \\
 15.01 & 3.80 & NS & Y \\
 16 & 4.16 & NS & Y \\
 17 & 1.95 & NS & Y \\
 18 & 4.23 & NS & Y \\
 18.5 & 3.85 & NS & Y \\
 19 & 4.05 & NS & Y \\
 19.56 & 3.86 & BH & Y \\
 20 & 3.84 & NS & Y \\
 21.68 & 1.57 & NS & Y \\
 23 & 4.20 & NS & Y \\
 24 & 3.82 & NS & Y \\
 25 & 3.83 & NS & Y \\
 40 & 1.62 & BH & Y \\
 60 & 4.45 & NS & Y \\ [1ex]
 \hline
\end{tabular}
\caption{Summary of the simulation models used here, by Burrows, Wang and Vartanyan (BWV) \cite{Burrows:2024pur}. Models are labeled by their progenitor masses (in units of solar mass, $M_{\odot}$). NS (BH) indicates progenitors whose compact remnants are neutron stars (black holes). Exploding progenitors are indicated with Y, while non-exploding progenitors are listed as N.\label{progentable}}
\label{table:1}
\end{table}

\begin{figure*}
\includegraphics[scale=0.65]{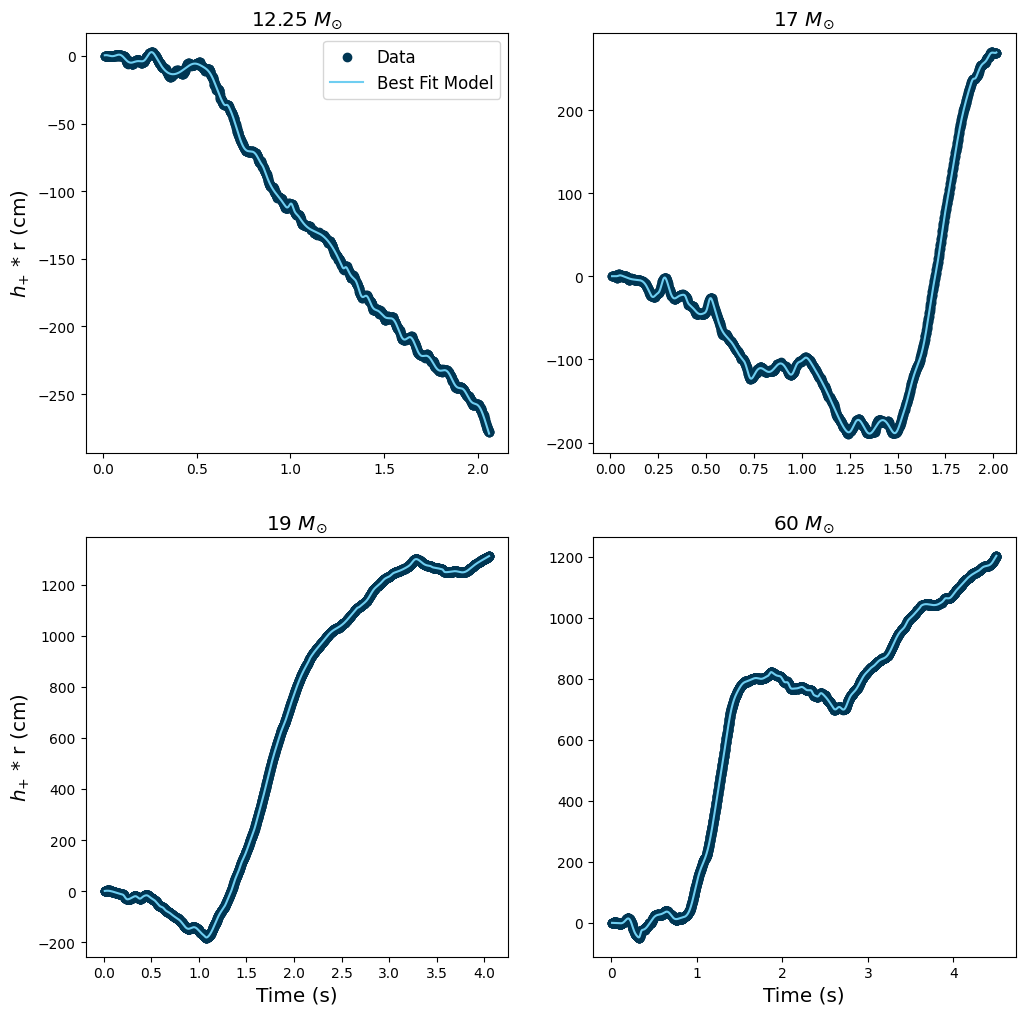}
\caption{\label{memplusfit} Fits to memory strain profiles, for the plus mode, for representative progenitor models.
The dimensionless strain has been scaled by the fiducial distance to the supernova, $r=10$ kpc. The dark blue points represent the strain profile from each simulation, while the cyan line represents our phenomenological fit.}
\end{figure*}

\section{ Modeling supernova GW waveforms }

\subsection{Simulations} \label{sims}

As the neutrino gravitational wave memory signal directly depends on anisotropies in neutrino emission, it is critical to fit our phenomenological model to three-dimensional simulations, as these are able to capture the full asymmetries of the supernova explosion and resulting neutrino emission. Additionally, it is important to choose simulations of multi-second duration to capture the full evolution of the memory signal.

We fit our phenomenological model to spectra from a suite of twenty state-of-the-art 3-D core-collapse supernova simulations performed by Burrows, Wang and Vartanyan (BWV from here on) \cite{Burrows:2024pur}. These simulations are based on solar-metallicity, non-rotating, spherical stellar progenitors from Sukhbold et al. \cite{Sukhbold:2015wba,Sukhbold:2017cnt}, and incorporate a variety of progenitor masses ranging from $9$ $M_{\odot}$ to $60$ $M_{\odot}$. While two simulations were provided for the $9$ $M_{\odot}$ progenitor - one including the effects of convection, one without - we chose to use only the model without convection, for consistency, as none of the larger progenitor simulations included convection effects. Simulation of radiation transport and hydrodynamics was carried out on each progenitor using the  \texttt{Fornax} code \cite{Skinner:2018iti}. Magnetic fields were not included. The BWV simulations extends to several (up to $\sim$4.5) seconds post-bounce, showing evidence of neutrino memory still developing after the accretion phase and well into the cooling phase, $t\gtrsim 1$ s. Table \ref{progentable} provides a summary of the simulations' inputs and results; see Ref. \cite{Vartanyan:2023sxm,Burrows:2024pur} for more details. 

\begin{figure*}
\includegraphics[scale=0.65]{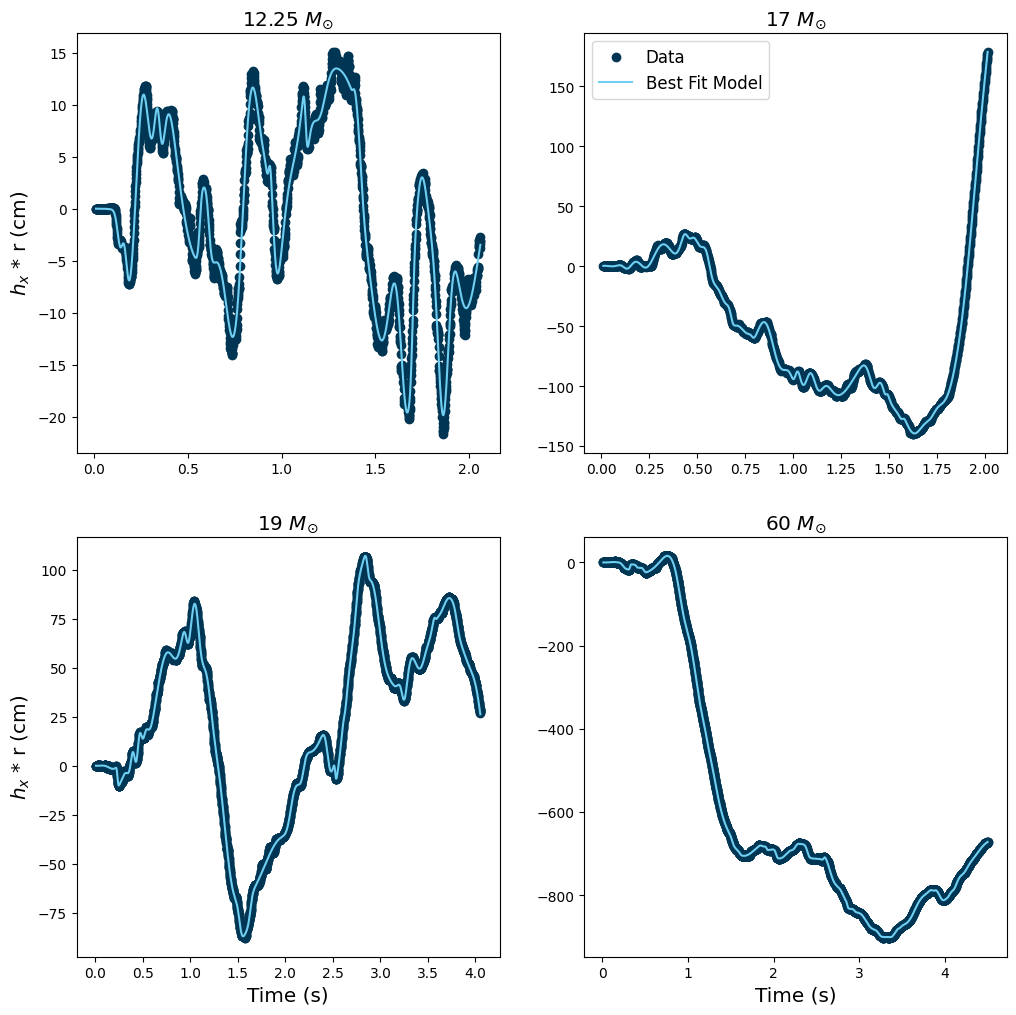}
\caption{\label{memcrossfit}Fits to memory strain profiles,  for the cross mode, for representative progenitor models. The same y-axis scaling convention and color profile is used as in Figure \ref{memplusfit}.}
\end{figure*}

The neutrino gravitational wave memory for all models is provided, separately for plus and cross modes, as the dimensionless strain multiplied by the distance to the supernova ($r=10$ kpc for all progenitors). Fits were performed directly to these scaled strains, and the amplitude parameters $\kappa_{i}$ were subsequently rescaled when calculating the contributions of each model to the SGWB. The matter contributions to the gravitational wave signal were provided as the amplitude spectral density, $S=2\sqrt{f}\text{ }h_{c}\left(f\right)$, where $h_{c}\left(f\right)$ is the characteristic strain,
\begin{equation}
h_{c}\left(f\right)=\sqrt{0.5\left(\left|\tilde{h}_{+}\left(f\right)\right|^{2}+\left|\tilde{h}_{\times}\left(f\right)\right|^{2}\right)}.
\end{equation}
Individual plus and cross mode contributions were not provided.

For each model, three viewing directions were available, with the observer located along the x-, y-, or z-direction with respect to the source. For this work, we choose the x-direction as our fiducial observer orientation; in Appendix \ref{inputVariation} we investigate the effects of averaging over the other viewing angles.

\subsection{Model Fitting}\label{modelfitting}

\begin{figure*}
\includegraphics[scale=0.65]{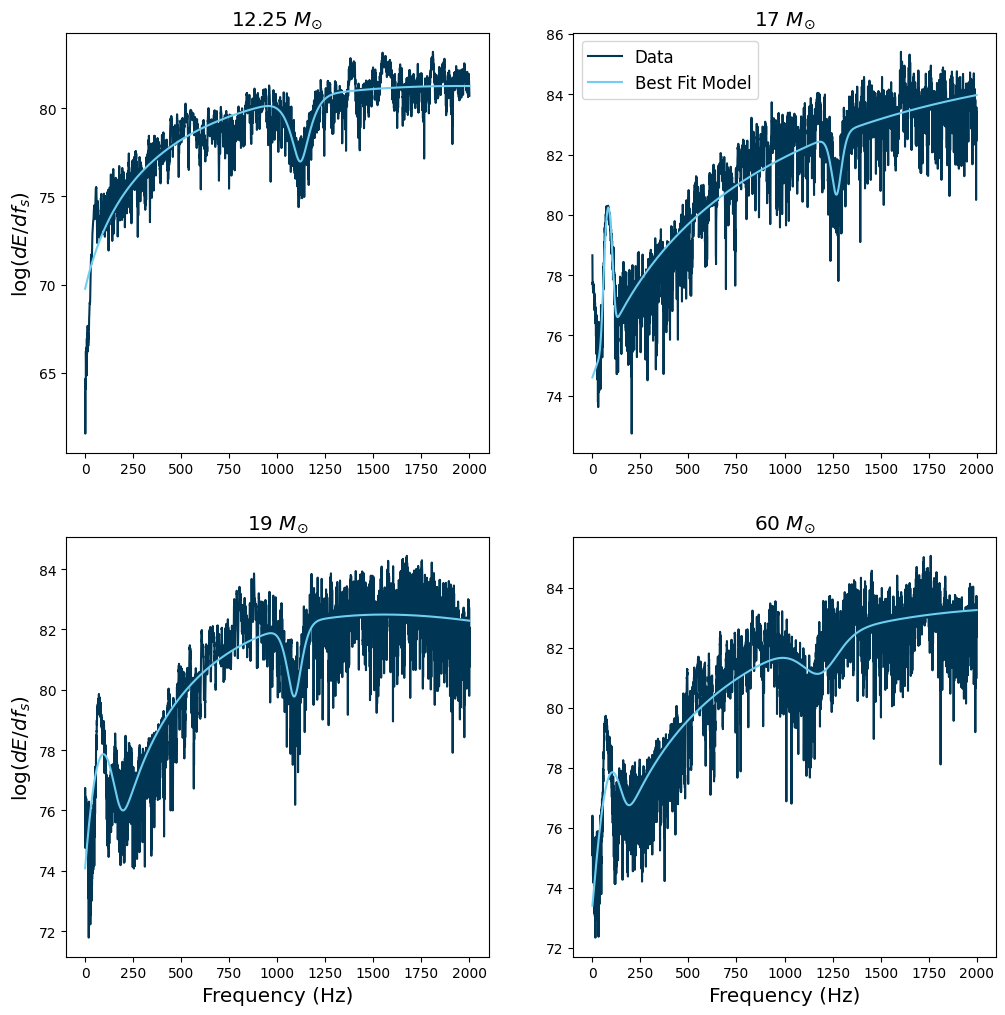}
\caption{\label{matterfits}Representative fits to the matter spectra. The dark blue line represents the natural logarithm of the matter contribution to $dE/df_s$, while the cyan line represents the phenomenological fit, Eq. (\ref{matterpheno}).}
\end{figure*}

We fit the numerical neutrino memory waveforms (precisely, the rescaled strains, $r h_\times(t)$ and $r h_+(t)$) to our phenomenological model, Eq. (\ref{nuphenostrain}). Since in the BWV simulations the neutrino memory evolves for several seconds after core bounce, the number of Gaussian terms in $\alpha_{i}$ (Eq. (\ref{nuanisotropy})) must be $\mathcal{O}\left(10\right)$ or more in order to accurately model the signal. To this end, instead of fixing $N$ we treat it as a parameter when fitting the phenomenological model to simulation data. As $N$ is \textit{a priori} unknown, for a given numerical data set, it is allowed to vary in the interval $N=[1,N_{max}]$, with $N_{max}=40$ chosen for computational reasons. For each (fixed) value of $N$, a least squares fit is performed, and the best-fitting values of the parameters $\left\{\kappa_{j},\zeta_{j},\tau_{j}\right\}$ are obtained. Then, the Bayesian Information Criterion (BIC) \cite{Schwarz:1978tpv,Neath:2012} is used to select the case (i.e., the value of $N$, $N_{best}$) for which the simulation data are best reproduced. We use the BIC for model selection because in the limit of large numbers of data points, it is a consistent criterion that, compared to other model selection criteria, favors more parsimonious models over non-minimal ones \cite{Neath:2012}.

When performing the fixed-$N$ fits, 
we apply physically motivated constraints on the domains of $\left\{\tau_{j}\right\}$ and $\left\{\zeta_{j}\right\}$. Specifically, we constrain the fitted values of $\left\{\tau_{j}\right\}$ to lie within the duration of the simulation. For the $\left\{\zeta_{j}\right\}$, we place upper and lower limits on the domain to avoid fitting features that are too small (and thus likely to be due to numerical jitter) or too large to be justified by the duration of the simulation. Since the memory strain is expected to vary over characteristic timescales of $\sim 0.1$ s or larger \cite{Mukhopadhyay:2021zbt}, corresponding to frequencies of $\sim 10$ Hz or lower, we chose an interval loosely centered around this latter value, $0.01  \leq \zeta_{j}\leq 100$ Hz.

Figures \ref{memplusfit} and \ref{memcrossfit} show the fit results
for $r h_+(t)$ and $r h_\times(t)$ respectively, for  representative simulations. We can see that, overall, the quality of the fit is good: the phenomenological models with fitted parameters capture the main large-scale features in the strain profiles for each progenitor.  We stress that these large-scale features are the most important, because they are responsible for the peak of the memory contribution to the \sngwb, which is in the Hz and sub-Hz frequency range (see Sec. \ref{subsec:results}).

For the matter contribution, we used the characteristic strain of the matter contribution as detailed in Section \ref{formalism}, calculated $dE/df_s$ using Eq. (\ref{dedf}), then fit this quantity to the phenomenological model, Eq. (\ref{matterpheno}). For numerical reasons, these fits were performed in log space. Initial values for each parameter in the models were chosen based on visual inspection of the spectra. Figure \ref{matterfits} shows representative fits to a selection of progenitors. The fits appear qualitatively acceptable, capturing the key features of the spectra. Additional features appear in some models at $f \gtrsim 1250$ Hz, however their amplitudes are smaller than the amplitude of the feature associated with the avoided \textit{f-g} mode crossing, which is in turn subdominant to the contribution of PNS oscillations (see Secs. \ref{singlemodelest} and \ref{subsec:results}). Therefore, we do not explore these additional high-frequency features further.

\section{\label{sec:level2} From individual waveforms to the stochastic background}

\subsection{Single-model estimates}\label{singlemodelest}

\begin{figure*}
\includegraphics[scale=0.65]{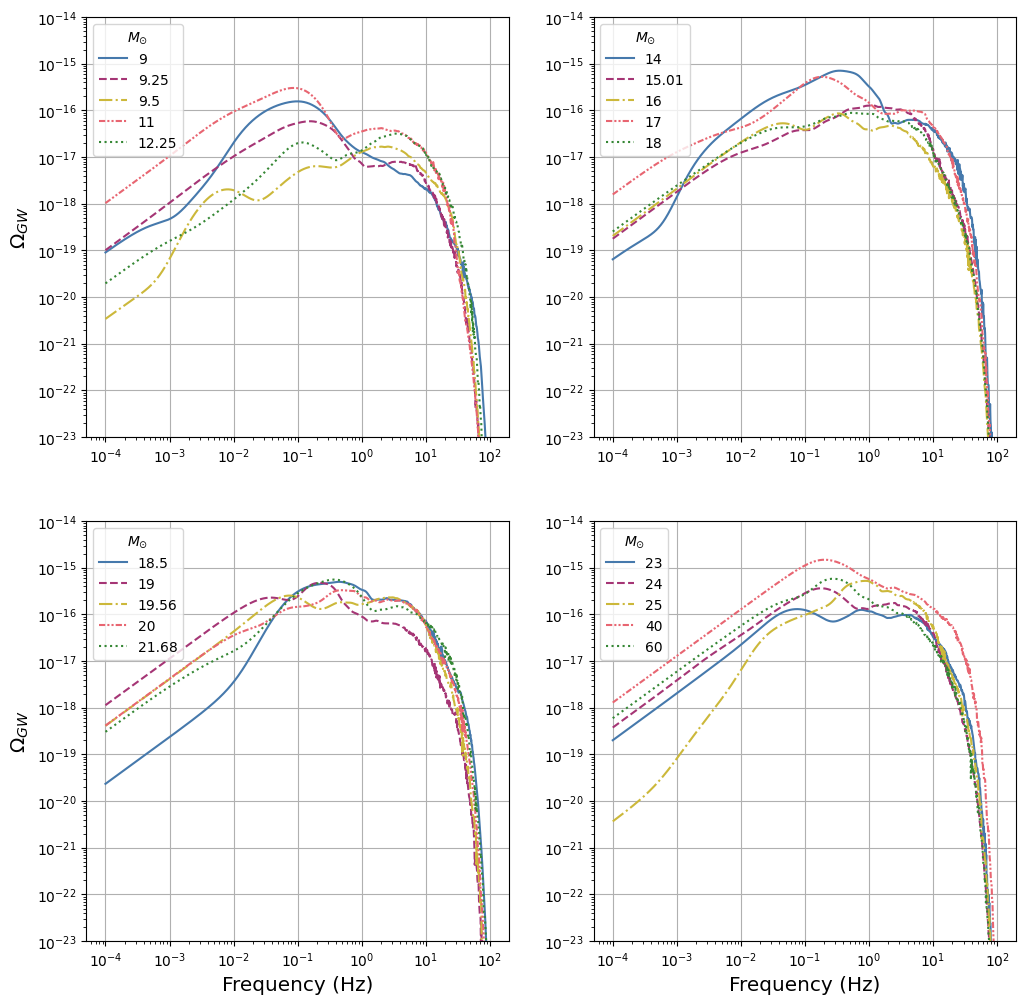}
\caption{\label{memoryfitcurves} Single-progenitor dimensionless energy densities for the neutrino memory component of the \sngwb, as a function of frequency, for the x-direction. Models are labeled by the progenitor mass, see Tab. \ref{table:1}.}
\end{figure*}

\begin{figure*}
\includegraphics[scale=0.65]{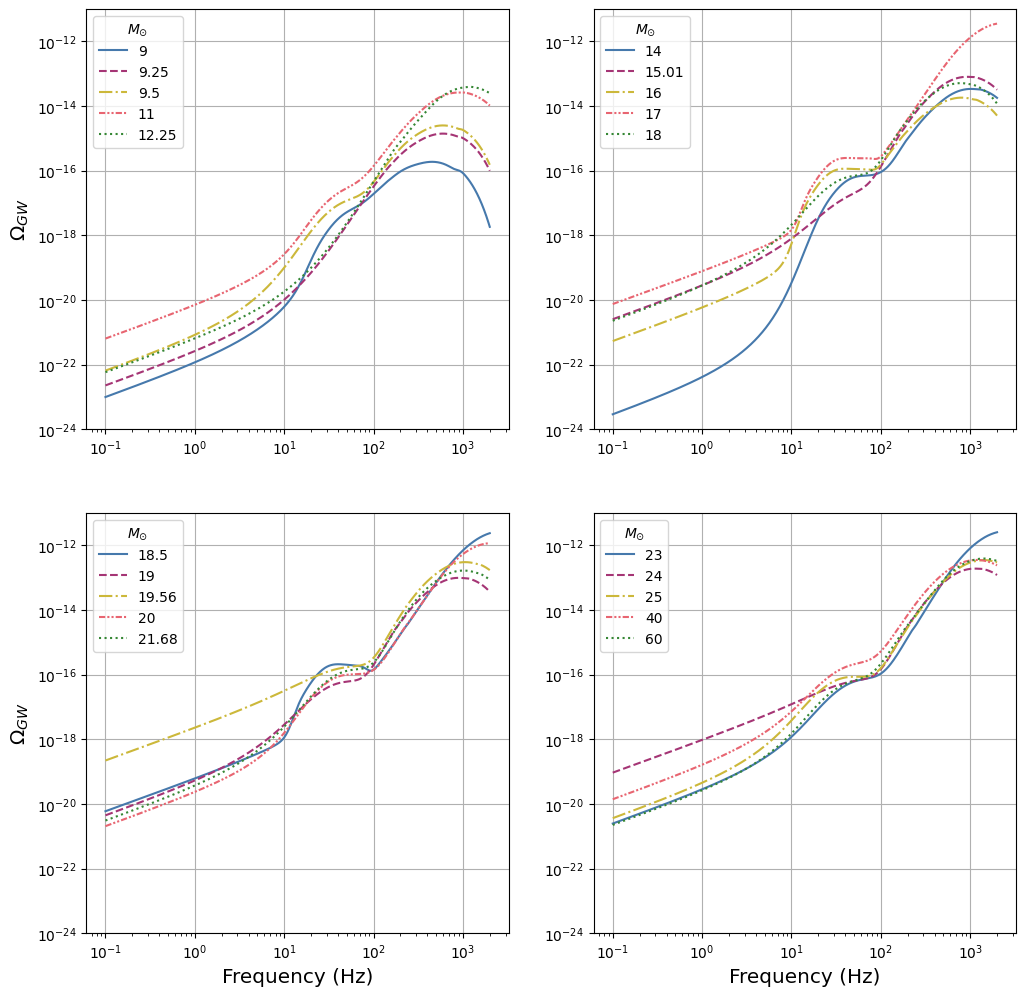}
\caption{\label{matterfitcurves} Single-progenitor dimensionless energy densities for the matter component of the \sngwb, as a function of frequency, for the x-direction. The same labeling scheme is used as in Figure \ref{memoryfitcurves}. }
\end{figure*}

As a step toward producing a realistic description of the \sngwb\, we first study single-model estimates, where the waveform for a single progenitor  is applied to entire stellar population, and used to compute the entire stochastic background (Eq. (\ref{eq:masterintegral})). This exercise is useful to illustrate how the features of a single waveform map into the spectrum of the stochastic background. 

Figures \ref{memoryfitcurves} and \ref{matterfitcurves} show the dimensionless energy density per logarithmic frequency interval, $\Omega_{GW}$, for each of these single-model estimates, for the memory and matter contributions, respectively. For the memory contribution, we see a mild general trend of increasing $\Omega_{GW}$ with increasing progenitor mass. This behavior is natural, as larger progenitors are more likely to undergo sustained turbulent accretion resulting in higher neutrino luminosities and anisotropies than smaller progenitors \cite{Burrows:2020qrp,Vartanyan:2023sxm}. Despite the general trend, however, we observe that the $M-\Omega_{GW}$ connection is not monotonic. Stochastic variation in these dynamics between ``neighboring'' mass progenitors (differing in mass by $\sim 1\msun$ or less )  -- especially when compared at a single viewing angle -- may be one contributing factor to the non-monotonicity. Additionally, the non-monotonic dependence on $M$ may be explained by the dependence on other key parameters. One of these is the core compactness (as defined in Ref. \cite{OConnor:2010moj}), which is strongly correlated with explosion energy \cite{Vartanyan:2023sxm}. We leave investigation of the relationship between neutrino memory strain and compactness to future work.

Despite these variations between individual models,  we can identify several general features that characterize the \sngwb\ from the neutrino memory: 
\begin{enumerate}
\item 
the presence of a low-frequency tail of power-law form, $\Omega_{GW}\propto f$, which is common to all models at $f\lesssim 10^{-3}~{\mathrm Hz}$ (extending to higher frequency for some models). This feature is due to 
the well known low frequency limit of the memory waveform, $\tilde h \propto 1/f$ for $f \rightarrow 0$ (see Eq. (\ref{numemspectrum})), which corresponds to $dE/df_s$ (Eq. (\ref{dedf})) tending to a constant; we refer to, e.g. \cite{Smarr:1977fy,Turner:1978jj,Epstein:1978dv,Buonanno:2004tp,Mukhopadhyay:2021zbt}, for details on this limiting behavior.

\item
A decihertz ($f\sim 0.1$ Hz) peak, which is expected considering the multi-second time scale evolution of the memory waveform ($\tau \sim {\mathcal O}(10)$ s, leading to $f \sim 1/\tau \sim {\mathcal O}(10^{-1})$ Hz). We observe a large degree of heterogeneity among models in the amplitude and width of this feature, and that several models exhibit multiple peaks in the $f \sim 0.01 - 0.1$ Hz range. The high degree of variability in the number and location of peaks in the frequency range may be related to the incomplete capturing of the multi-second GW memory evolution for most models, as this introduces an irreducible uncertainty into the fits for these features. Further, as the GW memory is characterized by anti-beaming, the observed signal amplitude is higher at observer angles orthogonal to the directions of highest neutrino luminosity \cite{Sago:2004pn,Leiderschneider:2021iah,Sakai:2025lks}. Differences in the signal amplitude between models may arise in part from misalignment between the observer and the axis along which the memory signal amplitude is highest. We interpret these structures further in Appendix \ref{inputVariation}, where we discuss the variability of the decihertz peak with respect to different observer viewing angles.

\item
A second, lower-amplitude peak in the $f \sim 1-10$ Hz band. This peak is present across all progenitors, and may reflect the fast variations (timescale $\tau \sim {\mathcal O}(1-10)$ ms) of the \n\ luminosities and  emission anisotropies due to hydrodynamical instabilities \cite{Mueller:2012sv,Mukhopadhyay:2021zbt,Muller:2026ofz}, such as strong post-shock convection and asymmetric accretion onto the PNS \cite{Lella:2026aat}. These variations may be primarily driven by rapid changes in the luminosity and anisotropy of $\nue$ and $\barnue$ during the accretion phase \cite{Richardson:2025ldi,Lella:2026aat}. While this component is also expected to exhibit some degree of dependence on observer orientation, we refer to Appendix \ref{inputVariation} for a more in-depth discussion.

\item 
A fast drop-off of $\Omega_{GW}$ at $f\gtrsim 10$ Hz. 
This trend is expected, considering that the memory effect is inherently a low-frequency phenomenon. 
It may partially enhanced by our fitting procedure (Section \ref{modelfitting}), where high-frequency features beyond a threshold were deliberately excluded. 
We anticipate that, in this high frequency region, the total \sngwb\ is dominated by the matter contribution (to be discussed below), therefore the fact that we potentially underestimate the memory part is unlikely to have a measurable impact. 

\end{enumerate}

Let us now discuss the matter contribution to the \sngwb\ in the single-model framework, as shown in Fig. \ref{matterfitcurves}.
Similar to the memory portion, the matter component from each model shows substantial heterogeneity. All models exhibit a peak at $\mathcal{O}\left(10^{3}\right)$ Hz, which is natural because this is the characteristic frequency of PNS oscillation modes. There is a general trend of increasing peak amplitude with increasing $M$, however the dependence is not monotonic. There is also a low-frequency tail of power-law form $\Omega_{GW}\propto f$, representing the zero frequency limit of the matter contribution. This is the matter memory contribution, and as it is much smaller than the neutrino memory component, we do not investigate it further.

Most models have a second peak at $\mathcal{O}\left(10^{1}\right)$ Hz, which might be due  to hydrodynamical instabilities originating from acoustic waves produced by prompt convection after core bounce \cite{Mueller:2012sv,Muller:2026ofz}. 
Exceptions are the $9.25$ and $12.25$ $M_{\odot}$ models, which lack the $\mathcal{O}\left(10^{1}\right)$ Hz peak. While the suppression due to an avoided f/g mode crossing (see Sec. \ref{sec:level2}) is visible in some models (e.g., the $9$ $M_{\odot}$ model), in most cases it is not a dominant feature. 

\subsection{ Population-averaged stochastic background}
\label{subsec:results}

\subsubsection{Fiducial Model}

\begin{figure*}
\includegraphics[scale=0.8]{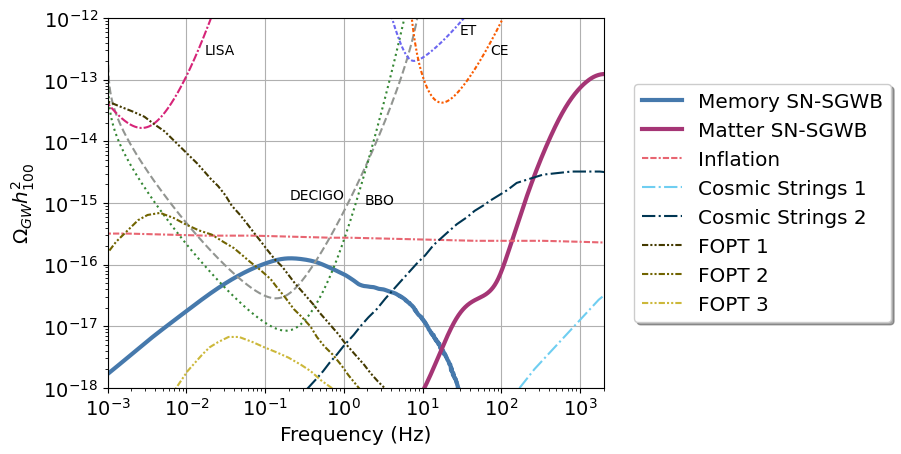}
\caption{\label{mainwithsens}Gravitational wave energy density of the \sngwb. The memory and matter contributions are shown separately, see legend. Five relevant detector power-law integrated sensitivity curves (PLISCs) are also shown (labels on curves), along with selected representative cosmological \gwb\ signals in the relevant frequency band: first-order phase transitions \protect\cite{Dunsky:2023ucb}, cosmic strings \protect\cite{Blanco-Pillado:2017rnf}, and inflation \protect\cite{Renzini:2022alw} (maximum allowed, see text for details).}
\end{figure*}

We now generalize the \sngwb\ calculation in Eq. (\ref{eq:masterintegral}) by including the waveform dependence on the progenitor mass, $M$, and integrating over the stellar population using the Salpeter IMF (see Sec. \ref{formalism}): 
\begin{equation}
\Omega_{\text{GW}}
\left(f\right)
=\frac{f}{\rho_{c}H_{0}}\int_{8 M_{\odot}}^{100 M_{\odot}} dM\text{ }w\left(M\right) \Omega\left(f,M\right)\label{eq:imfweightedintegral},
\end{equation}
where $\Omega(f,M)$ is the expression in Eq. (\ref{eq:masterintegral}) for a given model's waveform, and the weight factor $w\left(M\right)$ is
\begin{equation}
w\left(M\right)=\frac{\phi\left(M\right)}{\int_{8 M_{\odot}}^{100 M_{\odot}} dM\phi\left(M\right)}.
\end{equation}
As we have a finite number of waveforms, the integral over $dM$ is approximated by a sum over mass bins, where the edges of each bin were taken to be halfway between the two neighboring models' masses.

This population-averaged \sngwb\ is our fiducial model, and is shown in Fig. \ref{mainwithsens}.  The figure includes the contributions from neutrino memory and from matter effects, with  detector sensitivity curves plotted for comparison (see Sec. \ref{SNRobs} for a discussion of detectability). The dimensionless energy density from neutrino memory reaches $\Omega_{GW}\sim 10^{-16}$ at the peak, which is indeed close to the value we found in most single-model estimates (Sec. \ref{singlemodelest}). 
In Fig. \ref{mainwithsens} we see the main spectral features already noted for the single-model estimates; namely, the primary and secondary peaks at $f\sim 0.1$ Hz and $f\sim 1$ Hz respectively, and the tails at high and low frequencies, the latter consistent with the power-law dependence discussed previously (Sec. \ref{singlemodelest}). While small fluctuations from the main features are visible in the range $f\sim \left(10^{-1}-20\right)$ Hz, these are likely a consequence of the discreteness of our binned description, and should not be over-interpreted. 

For the matter contribution, we find a similar correspondence with single model estimates, with the  primary peak at $f \sim 1000$ Hz, the secondary peak near $f\sim 30$ Hz, and the power-law tail at low $f$. The peak value is $\Omega_{GW}\sim 10^{-13}$. This is much larger than the magnitude for single progenitors with $M\lesssim 16\msun$ or so (see Fig. \ref{matterfitcurves}), which dominate the rate of core collapse. Therefore, we conclude that the the matter part of  $\Omega_{GW}$ receives a substantial contribution from the high mass progenitors, $M\gtrsim 17\msun$, which are relatively rare ($\sim 30\%$ or so of all collapses) but are more powerful GW emitters. This suggests that the different contributions to the \sngwb\ spectrum may be sensitive to different segments of the stellar population. We note that the suppression due to the avoided f/g mode crossing is not evident in the waveform-averaged spectrum. This is expected given that this was a minor feature present only in the smallest-mass single-model spectra.

We emphasize that, while the main features described above are qualitatively robust, there are several sources of uncertainty in our fiducial model that can affect its normalization and the detail of its spectrum. Some of these are: the use of single waveforms to represent all stellar progenitors in a given mass band, the effect of observer orientation with respect to the source, uncertainties on the IMF, on the normalization and $z$-dependence of  the cosmic SFR, as well as on the fraction of black hole-forming and non-exploding progenitors. As a partial attempt to quantify these uncertainties, we explored the effect of varying certain parameters and assumptions. For all variations considered, the changes compared to our fiducial model were minimal. We present a summary of these findings in Appendix \ref{inputVariation}.

\subsection{Observational Prospects} \label{SNRobs}

The observational prospects for the \sngwb\ differ substantially for the matter and memory contributions. The matter contribution to the \sngwb\ is in principle detectable by both space-based and terrestrial detectors. On the other hand, the detection of the memory contribution has specific issues due to nature of the memory. As they employ free-falling masses, space-based detectors (e.g. LISA, DECIGO, BBO) are best-positioned to detect GW memory \cite{Favata:2010zu}. In contrast, terrestrial detectors are not inherently capable of maintaining the permanent displacement that is a signature of GW memory \cite{Favata:2010zu,Richardson:2025ldi}, but, for an individual supernova, they are sensitive to the low-frequency ramp-up signal leading up to the final plateau \cite{Richardson:2024wpp}. For this reason, we will include both space-based and terrestrial detectors in our discussion. 

The detection of a \gwb\ signal is accomplished through (i) cross-correlation of signals from two or more detectors, or (ii) auto-correlation of signals from a single detector, which is possible in the case of excellent noise subtraction \cite{Tinto:2001ii,Romano:2016dpx}. For a pair of detectors $I$ and $J$, the detectability of such a signal is expressed through the Signal-to-Noise Ratio (SNR) $\varrho_{IJ}$ \cite{Allen:1997ad}:
\begin{equation}
\varrho_{IJ}=\left[n_{\text{det}}t_{\text{obs}}\int_{f_{\text{min}}}^{f_{\text{max}}}df\text{ }\frac{\Gamma^2_{IJ}\left(f\right)\Omega^{2}_{\text{signal}}\left(f\right)}{D_{I}^{\text{noise}}\left(f\right)D_{J}^{\text{noise}}\left(f\right)}\right]^{1/2},
\label{eq:signaltonoise}
\end{equation}
where $D_{i}^{\text{noise}}\left(f\right)$ is the noise auto power spectrum of detector $i\in I,J$, $\Gamma_{IJ}\left(f\right)$ is the overlap reduction function that characterizes the effects of detector geometry and relative orientations, $t_{\text{obs}}$ is the total observation time, and $n_{\text{det}}$ is the number of detectors: $n_{\text{det}}=2$ in the case of cross-correlation between two different detectors ($I\neq J$), and $n_{\text{det}}=1$ in the case of auto-correlation within a single detector ($I = J$). 

As their peak amplitudes occur in different frequency bands, we perform separate auto-correlation ($n_{\text{det}}=1$) SNR calculations for the memory contribution and the matter contribution. For the memory part, we calculate the SNR for $t_{\text{obs}}=1$ year for DECIGO and BBO, as well as an improved version of DECIGO better equipped for removal of foreground contamination from neutron star binaries \cite{Yagi:2011wg,Kuroyanagi:2014qza}. For the matter contribution, we calculate the SNR for Cosmic Explorer (CE) and the Einstein Telescope (ET). In each SNR calculation, we use the functions and parameters for each detector as provided in Ref. \cite{Schmitz:2020syl}, Appendix A, and references therein. The resulting SNR values for each case are provided in Table \ref{table:2}. We find that, in the absence of other backgrounds, the \sngwb\ signal from memory is potentially observable by both DECIGO and BBO ($\varrho_{II} \gtrsim 1$). However, the signal from the matter contribution is  below the sensitivity of CE or ET by several orders of magnitude: we obtain $\varrho_{II} \sim {\mathcal O}(10^{-29})$ for the SNR (not shown in Table \ref{table:2}).

\begin{table}[h!]
\centering
\begin{tabular}{|c c|} 
 \hline
 Detector & SNR  \\ [1.0ex] 
 \hline
 DECIGO & 1.4 \\
 Improved DECIGO & 3.1 \\
 BBO & 10.3 \\ [1ex] 
 \hline
\end{tabular}
\caption{Calculated SNR values for the memory contribution to the \sngwb\, for three representative detectors with $t_{obs}=1$ yr. For the matter part and realistic detectors, the SNR is extremely small (see text) and is not included here.}
\label{table:2}
\end{table}

As a complement to the SNR, we study the detectability as a function of frequency by showing the Power-Law Integrated Sensitivity Curves (PLISCs) \cite{Thrane:2013oya,Schmitz:2020syl} for the above-mentioned detectors, see Figure \ref{mainwithsens}. PLISCs are constructed under the assumption that the {\gwb} signal power spectrum can be modeled by a power-law, taking advantage of the broadband nature of the {\gwb} signal to improve the SNR through an integration over frequency. Even if it does not strictly have a power-law form, a (potential) signal is considered detectable in the frequency interval where its dimensionless energy density exceeds the detector PLISC. We see that for both DECIGO and BBO, the \sngwb\ from neutrino memory is potentially detectable in a frequency interval of about 1-2 orders of magnitude centered at $f\sim 10^{-1}$ Hz \footnote{
We caution that the PLISC curves shown in fig. \ref{mainwithsens} are not directly related to the SNR values in Table \ref{table:2}. SNRs can be obtained from PLISCs only for signal spectra that are close to a power-law in the frequency band most relevant to each detector (see Ref. \cite{Schmitz:2020syl}), which is not the case here. }.

In addition to experimental capability and detector noise, the observability of the \sngwb\ depends on other competing signals that may overlap in frequency or amplitude. We first examine those signals that could be of comparable amplitude, which are particularly interesting because of the potential confusion between these and the \sngwb\. These competing signals could originate from first-order phase transitions \cite{Caprini:2015zlo,Schmitz:2020syl,Dunsky:2023ucb,Croon:2024mde,LIGOScientific:2025kry}, cosmic strings \cite{Blanco-Pillado:2017rnf,Auclair:2019wcv,Avgoustidis:2025svu}, or inflation \cite{Turner:1996ck,Barnaby:2011qe,Guzzetti:2016mkm,Bartolo:2016ami}. Figure \ref{mainwithsens} shows several first-order phase transition models from Ref. \cite{Dunsky:2023ucb}, two cosmic string models from Ref. \cite{Blanco-Pillado:2017rnf}, and the slow-roll inflation model from Ref. \cite{Renzini:2022alw}, which represents the maximum allowed by CMB data (tensor-to-scalar ratio $r_{TS} = 0.11$, which is the upper limit from \textit{Planck} \cite{Planck:2015fie}).  These models were chosen to illustrate the potential confusion with the \sngwb\. Considering their large uncertainties, the stochastic backgrounds from these sources could either greatly exceed the \sngwb\ or be overwhelmed by it. Therefore, if these cosmological signals remain undiscovered as detectors reach the sensitivity of $\Omega_{GW}\sim 10^{-16}$, it will become increasingly important to take the \sngwb\ into consideration. 

Searching for \emph{any} signal at the $\Omega_{GW}\sim 10^{-16}$ level is in itself a formidable challenge. 
The main obstacle is the overwhelming background from CBCs, which can be as high as $\Omega_{GW}\sim 10^{-12}$-$10^{-11}$ at $f\sim 0.1$ Hz, and $\Omega_{GW}\sim 10^{-9}$ at its peak frequency, $f\sim 10^3$ Hz (see, e.g., \cite{Zhu:2012xw,Bellie:2023jlq} for the theory, and \cite{KAGRA:2021kbb} for observational upper bounds). It is clear that a very powerful subtraction of this foreground, with a reduction factor of at least 4 orders of magnitude, would be needed to achieve sensitivity to the \sngwb\ and other signals comparable to it. We discuss possible mitigation strategies in Section \ref{sec:discussion}.

\section{\label{sec:discussion}Discussion} 

We have modeled the supernova stochastic gravitational wave background (\sngwb), in its two major components: GWs produced by matter contributions, and the GW memory from asymmetric neutrino emission. The two components were modeled consistently, using the results of the same suite of state-of-the-art, three-dimensional, multi-second core collapse simulations \cite{Burrows:2024pur}.
Our estimate for the matter contribution is broadly consistent with  the  literature (e.g., \cite{Finkel:2021zgf,Chowdhury:2024fdr}), and shows the characteristic peak at $f\sim 10^{3}$ Hz. We find that the neutrino memory component occupies a distinct frequency band, $f\sim 10^{-2}-1$ Hz, and therefore may be distinguishable from the matter component. It reaches a peak value of $\Omega_{GW}\sim 10^{-16}$ at $f\sim 10^{-1}$ Hz.  Despite the various uncertainties at play, our results are reasonably robust with respect to variations in source orientation and in core collapse rate. While the matter contribution is well below experimental sensitivity, the memory component offers the best prospects for observation: we find that it is within the reach of planned next-generation decihertz detectors like DECIGO and BBO. Under ideal circumstances, where foregrounds are completely subtracted (see below), the $f\sim 0.1$ Hz peak might be identified above the noise (signal-to-noise ratio $\gtrsim$ 1) in as little as one year of data-taking. 

If detected, the memory component of the {\sngwb} could provide population-level information about supernovae and their progenitors that is complementary to existing data from other sources. From the memory signal, one could potentially extract information about the average anisotropy and luminosity of neutrino emission in supernova explosions, as well as the fraction of black hole-forming supernovae. Additionally, the {\sngwb} as a whole is complementary to the Diffuse Supernova Neutrino Background (DSNB), as both originate from the same source population. While the DSNB is likely to be detected first, given recent excess seen at Super-Kamiokande \cite{Super-Kamiokande:2021jaq,Super-Kamiokande:2025sxh}, such a detection could further inform predictions of the {\sngwb}. We leave further discussion of the complementarity of the {\sngwb} and the DSNB to future work. Finally, detection of the memory component of the {\sngwb} would confirm a prediction of general relativity that has yet to be measured.

We note that our estimate of the \sngwb\ is subject to several limitations. Currently, no 3D core-collapse supernova simulations exist that are of adequate duration to capture the full evolution of the neutrino memory signal. In light of this, our estimate for the neutrino memory component of the \sngwb\ may be considered conservative. Additionally, our use of single waveforms to represent each mass band may introduce spectral features that arise not from general characteristics of the \sngwb\, but from the stochastic differences between models. Four of the models used -- the $12.25$ $M_{\odot}$, $14$ $M_{\odot}$, $19.56$ $M_{\odot}$, and $40$ $M_{\odot}$ models -- resulted in the formation of a black hole rather than a neutron star, and were associated with qualitatively different gravitational wave signals that the neutron star formers \cite{Choi:2024irp}. However, numerous simulation results in the literature \cite{OConnor:2010moj,Pejcha:2014wda,Sukhbold:2015wba,Muller:2016ujh,Ertl:2015rga} report ``bands'' of black hole formation associated with failed supernovae interspersed with successful explosions. Possible physical explanations for this pattern include the effects of metallicity \cite{OConnor:2010moj,Pejcha:2014wda}, progenitor rotation \cite{OConnor:2010moj}, the PNS equation of state  \cite{Pejcha:2014wda}, pre-supernova progenitor structure \cite{Sukhbold:2015wba}, and fallback \cite{Muller:2016ujh}. For this reason, we consider the inclusion of black hole-forming models as an acceptable approximation  given the limitations of our simulations. Finally, we note that our estimate of the \sngwb\ depends on the assumptions underlying the simulations we have used, and may be different if another set of simulations were to be used. For example, our estimate does not include effects arising from rotation or magnetic fields as these were not included in the simulations \cite{Vartanyan:2023sxm, Choi:2024irp}. As new state-of-the-art simulations including these effects become available, our model could be adapted to include them. In particular, we anticipate that including the effects of rotation may result in a mild enhancement of our predicted signal \cite{Pajkos:2025zob}. Our results represent a step in the direction of more realistic modeling of the \sngwb.

We found that, near its peak, the \sngwb\ is within a factor of a few from the
current upper limit for the 
\gwb\ from slow-roll inflation, which lies at $\Omega_{GW}\sim 3~10^{-16}$. Stochastic gravitational waves from inflation are a target of prime interest for cosmology, and therefore reaching a sensitivity of $\Omega_{GW}\sim {\mathcal O}(10^{-16})$ is an important goal of future GW detectors \cite{Corbin:2005ny,Sedda:2019uro}. While indirect detection of the inflationary \gwb\ through measurements of primordial B-modes in the CMB may occur in the near future, a direct detection of the \gwb\ could provide a wealth of complementary information \cite{Maggiore:2018aaa}. The landscape of GWs at or near that scale could be even richer, including possible contributions from first order phase transitions and cosmological relics. Once the relevant sensitivity is reached, the problem of disambiguation between signals from different classes of sources will have to be addressed. 
It is possible that, in the search for cosmological GWs, the neutrino memory component of the \sngwb\ will play the role of a gravitational ``neutrino fog", similar to the neutrino fog that limits Dark Matter searches at cryogenic detectors. We found that the \sngwb\ has a minimum 
at $f\sim 10-30$ Hz, which might therefore be optimal for searching for stochastic backgrounds from other sources (see fig. \ref{mainwithsens}).

Detecting stochastic backgrounds at the level of $\Omega_{GW}\sim {\mathcal O}(10^{-16})$ will require 
subtracting the overwhelming background from CBCs, which dominates over most other \gwb\ signals in the frequency bands at and around its peak amplitude \cite{LIGOScientific:2017zlf,Christensen:2018iqi, KAGRA:2021kbb,LIGOScientific:2021qlt,LISA:2022yao,Lehoucq:2025ruc}. While next-generation ground-based interferometers may reach sensitivities adequate to completely resolve (and therefore subtract) the background from binary black hole (BBH) CBCs, binary neutron star CBCs would remain difficult to fully resolve \cite{Regimbau:2016ike,Sachdev:2020bkk,Pan:2023naq}, and uncertainties in modeling and source parameters limit the ability to remove these backgrounds \cite{Zhou:2022nmt,Kume:2024xvh,Song:2024pnk}.

Recently, machine learning methods have been successfully used to detect and subtract the \gwb\ from BBH CBCs from a mock dataset representative of the LIGO-Virgo-KAGRA network at design sensitivity \cite{Einsle:2025xsh}. With further advances in this direction, it may be possible to reduce the CBC background enough to reveal subdominant signals, including the \sngwb\. If cosmological signals are constrained to amplitudes below the predicted \sngwb\, these methods could in theory provide an avenue by which the \sngwb\ could itself be identified and subtracted. This will enable studying the physics of the supernova neutrino memory, and extending the sensitivity to stochastic backgrounds below the  $\Omega_{GW}\sim {\mathcal O}(10^{-17}-10^{-16})$ scale.
For such an approach to be effective, it is critical that the \sngwb\ is modeled to an adequate degree of precision, as even small mischaracterizations could negatively impact the effectiveness of background subtraction. This is especially true for the case of deep neural networks, which often require large, high-quality training sets to perform well. 

To sum up, we have shown that the stochastic gravitational wave background from asymmetric neutrino emission (memory effect), is a relevant (potential) signal for next generation decihertz detectors. It represents a unique probe of core collapse physics, and could play an important role in the exciting search of gravitational waves from inflation and other cosmological phenomena. Our work provides a step toward fully realistic models of this background, which we hope will be instrumental for the scoping of future GW detectors.

\section*{Acknowledgments}
We are grateful to the authors of Ref. \cite{Burrows:2024pur}, for clarifications and insights on their work, and to John Beacom, Giuseppe Lucente, Mainak Mukhopadhyay and Garv Chauhan for useful discussions. We acknowledge support by the NSF Award Number 2309973.

\appendix

\section{Varying assumptions and  parameters}\label{inputVariation}

\begin{figure*}
\includegraphics[scale=0.65]{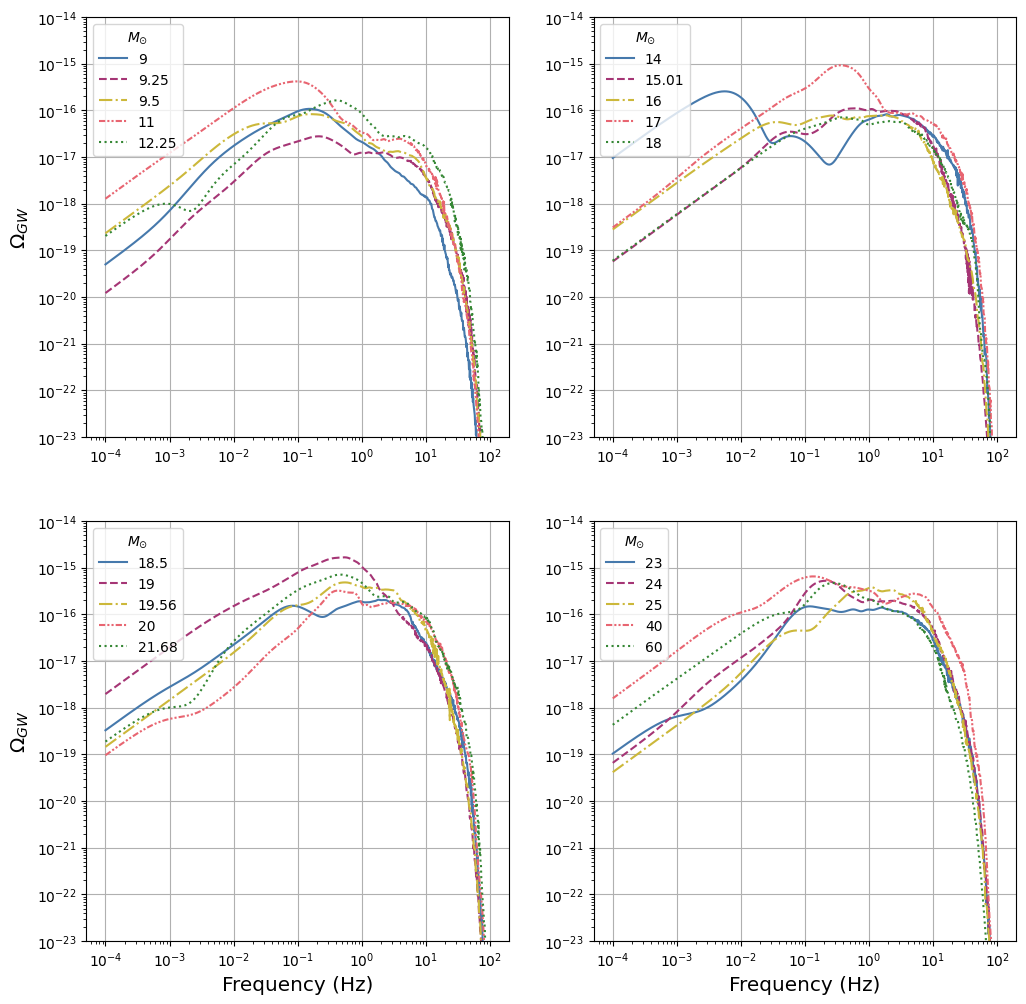}
\caption{\label{memoryfitcurvesY} Single-progenitor dimensionless energy densities for the neutrino memory component of the \sngwb, as a function of frequency, for the y-direction. The same color scheme is used as in Figure \ref{memoryfitcurves}.}
\end{figure*}

\begin{figure*}
\includegraphics[scale=0.65]{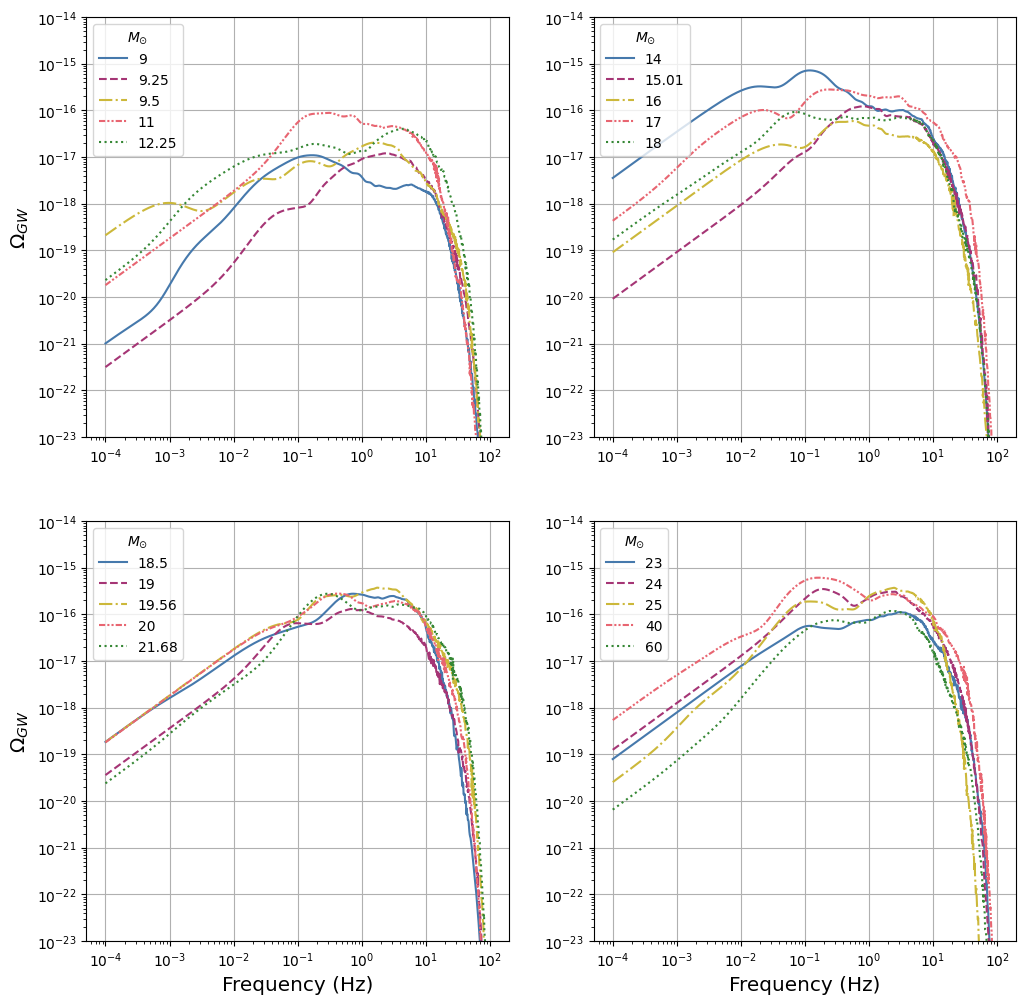}
\caption{\label{memoryfitcurvesZ} Single-progenitor dimensionless energy densities for the neutrino memory component of the \sngwb, as a function of frequency, for the z-direction. The same color scheme is used as in Figure \ref{memoryfitcurves}. }
\end{figure*}

\begin{figure*}
\includegraphics[scale=0.75]{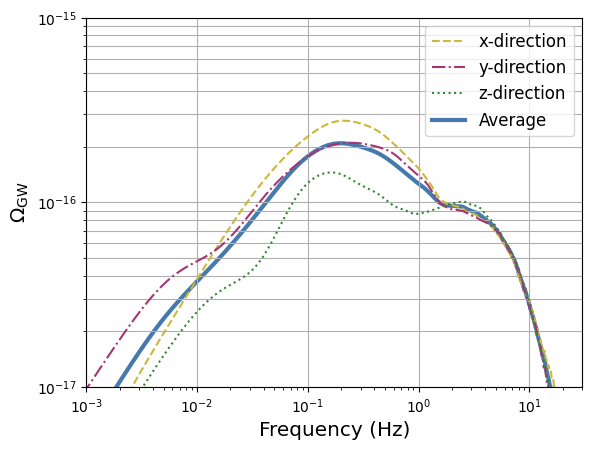}
\caption{\label{angavgsgwb}Gravitational wave energy density of the total \sngwb\, for the x-, y-, and z-direction source orientations and source orientation-averaged total, Eq. (\ref{eq:avgorientations}).}
\end{figure*}

In this Appendix, we estimate the degree of uncertainty on our fiducial model, Fig. \ref{mainwithsens}, by checking how variations in assumptions and parameters affect the results for $\Omega_{GW}$. First, we estimate the impact of using  different source orientations (with respect to the observer) on the memory contribution. We repeated our analysis using the memory waveforms in the y- and z-direction orientations. Figures \ref{memoryfitcurvesY} and \ref{memoryfitcurvesZ} show single-model \sngwb\ memory results for the y- and z-orientations, respectively. These should be compared with Fig. \ref{memoryfitcurves}, where the x-direction was used. For most of the progenitor models, the number, amplitude, and width of the $f \sim 0.01 - 0.1$ Hz features change considerably between different observer orientations. Models with a single dominant feature in this frequency band for one orientation often demonstrate a substantial reduction in amplitude of the same feature in the other orientations. This behavior is consistent with the interpretation of these features given in Section \ref{singlemodelest}.

In contrast, the $f \sim 1$ Hz peak shows much less variation in amplitude across orientations. This is consistent with our interpretation of these features as representing a superposition of short-timescale, stochastic changes in neutrino emission anisotropy arising from hydrodynamical instabilities, causing what are effectively small bursts of memory superimposed on a larger evolution trend. While individually these contributions should exhibit variations with orientation similar to the lower frequency structures, once considered together as a whole they exhibit similar distributions in frequency for all orientations. 

Using the results in Figs. \ref{memoryfitcurvesY} and \ref{memoryfitcurvesZ}, we have computed the population-averaged memory \sngwb\ for the three different (fixed) observer orientations (Eq. (\ref{eq:imfweightedintegral})). Furthermore, 
as a step toward a more realistic modeling, we have performed an average of the results for the three orientations, by computing
\begin{eqnarray}
\bar{\Omega}_{\text{GW}}
\left(f\right)
=&\frac{1}{3}\sum_{i\in x,y,z}\Omega_{\text{GW},i}
\left(f\right)~,
\label{eq:avgorientations}
\end{eqnarray}
where $i\in x,y,z$ refers to the axis along which the observer is oriented, and $\Omega_{\text{GW},i}$ refers to the \sngwb\ as calculated for each observer orientation using Eq. (\ref{eq:imfweightedintegral}).
The expression in Eq. (\ref{eq:avgorientations}) is a rough proxy of what might be obtained by a detailed integration over the angular orientations of a randomly distributed stellar population. 

Figure \ref{angavgsgwb} shows the memory contribution to the population-averaged \sngwb\ for the three different (fixed) observer orientations, along with the orientation-averaged spectrum, Eq. (\ref{eq:avgorientations}). We find that while the directional-dependence of the $f\sim 0.1$ Hz peak results in a smaller amplitude for this feature after averaging, the variation is less than an order of magnitude. The $f\sim 1$ Hz feature shows much less dependence on source orientation, as is expected based on the results from the single-waveform analysis. The qualitative behavior at very low and high frequencies remains the same between orientations. Therefore, the uncertainty due to the orientation direction is likely to be subdominant compared to  
uncertainties of other kinds. Interestingly, the averaged spectrum in Fig. \ref{angavgsgwb} is very close to the one obtained using the x-direction, therefore our fiducial model in Fig. \ref{mainwithsens} can be considered robust. 

For the matter contributions, all three source orientations produced frequency spectra that were similar in overall amplitude and shape at $f\gtrsim 1000$ Hz. Instead, the low-frequency ($f\sim 100$ Hz) feature was either reduced in amplitude or not present in the z-direction spectra.  This is expected to cause an $\mathcal{O}\left(1\right)$ or smaller change to this feature upon averaging over all three orientations. Given the poor prospects for detectability in this part of the frequency spectrum, we do not further pursue orientation-averaging for the matter contribution.

Next, we explore the impact of variations in the core collapse rate (CCR) given by Eq. (\ref{eq:CCR}). The uncertainty in the CCR has contributions from the SFR (Eq. (\ref{eq:SFR})) and from $\lambda_{CC}$ (see Sec. \ref{formalism}). As $\lambda_{CC}$ is estimated from the IMF, and as the cosmic average IMF for progenitor masses above $8$ $M_{\odot}$ is well-modeled by the Salpeter IMF used in Sec. \ref{formalism} \cite{Kroupa:2000iv,Chabrier:2003ki,Horiuchi:2011zz,Aoyama:2021ltt}, the dominant contribution to the uncertainty in Eq. (\ref{eq:CCR}) arises from the SFR. Recent results from JWST have suggested that the cosmic SFR may evolve differently with redshift than previously thought \cite{Harikane:2021rqt,Harikane:2022rqt}.

In an attempt to quantify the uncertainty introduced by the SFR model, we investigated the effect of different SFR model choices on our estimate for the neutrino GW memory component of the \sngwb\. We repeated the \sngwb\ calculations for the neutrino memory component using several SFR model variants from Refs. \cite{Madau:2014bja,Vangioni:2014axa,Harikane:2021rqt}. All SFR models led to variations of $\mathcal{O}\left(1\right)$ or less from the fiducial model despite the differences in behavior at high redshift. We therefore conclude that despite the uncertainty associated with the cosmic SFR at high redshifts, the choice of SFR model is unlikely to impact our predictions significantly.

\bibliography{references}

\end{document}